\documentstyle[aps,prl,twocolumn,floats,epsfig]{revtex}
\begin{document}
\wideabs{
\title{Self Interaction Correction with Unitary Invariance}
\author{Mark R Pederson$^1$, Adrienn Ruzsinszky$^2$ and John P. Perdew$^2$}
\date{\today}
\address{$^1$Department of Chemistry, Johns Hopkins University, Baltimore MD, 21218 and Office of
Basic Energy Sciences, SC22.1, US Department of Energy, Washington DC, 20585}
\address{$^2$Department of Physics, Temple University, Philadelphia PA 19122}
\date{\today}
\maketitle
\begin{abstract}
Standard spin-density functionals for the exchange-correlation
energy of a many-electron ground state make serious self-interaction
errors which can be corrected by the Perdew-Zunger self-interaction
correction (SIC). We propose a size-extensive construction of SIC orbitals which,
unlike earlier constructions, makes SIC computationally efficient,
and a true spin-density functional. The SIC orbitals
are constructed from a unitary transformation that is explicitly
dependent on the non-interacting one-particle density matrix.
When this SIC is applied to the local spin-density approximation, improvements are
found for the atomization energies of molecules.
                                                                                                                               
\end{abstract}
}
 
~\\
Density Functional Theory (DFT)~\cite{R2,R3} allows for a quantum-mechanical description of
electrons without direct calculation of the many-electron wavefunction.
Approximations to DFT (including the local-spin-density or LSD approximations~\cite{R3} and
gradient approximations~\cite{PW92,pbe}) are
widely and successfully used to predict, understand and optimize many physical and chemical phenomena. However, such
approximations have yet to be cast in a form that is both efficient and free
of the self-interaction error~\cite{R1}.
For example, in neutral systems the long-range behavior of the Kohn-Sham (KS)
potential does not reduce to the $-1/r$ form expected from general considerations.
As discussed in Refs.~\cite{R28,RLT}, the
incorrect asymptotic form of the effective potential leads to a range of related issues, sometimes
referred to as delocalization errors~\cite{aron}, when using
DFT functionals to understand chemistry, materials and physics.
                                                               
For open-shell systems, standard DFT expressions allow
for relatively accurate predictions of electron affinities if total energy differences are calculated.
However, because of the self-interaction error, the highest anionic KS eigenvalue is usually positive, suggesting that
an isolated anion would prefer to have fractional charge.
A more serious deficiency arises for closed shell
molecules.
While anions of high-symmetry molecules are often experimentally unstable, molecules
with large dipole moments, such as ethylene carbonate~\cite{dipole1}, are known to form dipole-bound
anions that are held together by electrostatics, dispersion and correlation~\cite{dipolebound}.
However, the SIC for the extra electron is large compared to these effects, which makes direct calculation of
weak interactions between excess electrons and dielectric media difficult within DFT.
In this regard, we note further that calculations of Rydberg-excitations in atoms,
or the Mott-Gurney continuum~\cite{mott} of shallow-defect states
in solids, would require the asymptotically correct forms of the effective potential that many
versions of SIC reproduce. Various constrained DFT methods~\cite{klein,dederich,baruah} have been suggested
to help address this problem for calculations of excited states.
Fractional dissociation of ionic molecules and crystals is another case where
most approximations to DFT, including hybrid DFT, present problems~\cite{dissociation}.
For such cases DFT-based methods
usually lead to improper dissociation with the ground-state separated limit leading to partially
charged atoms.
We mention parenthetically that in regard to bond-breaking of condensed phases into separated atoms
and broken-symmetry calculations that are often used for magnetic calculations~\cite{molmag},
there is the so-called symmetry dilemma due to the fact that a single determinant constructed
from the KS orbitals is not always an eigenstate of symmetry or spin; there are good physical reasons~\cite{dilemma} to permit
such symmetry breaking in density functional approaches including ours here.
Calculations of DFT atomic energies as a function of fractional occupation show, for at least localized
systems, that the energy as a function of charge state is not linear in the fractional electron limit as
required by the Perdew-Levy theorem that requires piece-wise linear energies and derivative discontinuities
at integer electron states~\cite{R28}.
With respect to modeling of electronic transport, simple atomic SIC methods have been shown to
be useful to more accurately account for the electronegativity differences between gold leads
and molecular islands~\cite{sanvito}.
An easy-to-use
self-interaction correction is very much needed in time-dependent DFT~\cite{tddft,hoffmann}, where
ionizations and atomizations occur continuously in real time.
While current implementations of SIC have not provided empirical evidence that the SIC error is the accuracy
limiting issue for calculation of atomization energy in molecules~\cite{ernzerhof,vydrov},
the need for accurate
reaction energies and reaction barriers has historically been one of the strong drivers for the development
of accurate approximations to DFT. It may be that a more systematic
recipe for incorporating SIC into DFT
will eventually provide an additional means for improving our ability to predict energetics
related to chemical transformations.
                                    

Other manifestations of this issue impact predictions of magnetic properties. As originally pointed out by
Janak {\em et al}~\cite{janak}, an incorrect accounting for the self-interaction correction can lead to
incorrect valences for open-shell atoms with partially filled $d$- and $f$- shells. Partially occupied
valences are not necessarily a physical issue as it is possible to obtain such interpretations from the
natural bond-orbital
analysis of multi-configurational wave functions and density matrices.
However,
Svane and Temmerman and coworkers have found that the calculation of phase diagrams in
systems containing $f$ electrons requires inclusion of SIC because the number
of occupied $f$ electrons changes as a function of lattice constant~\cite{temmerman}.
In systems composed of lighter elements, such as metal-oxide-based molecular magnets, DFT often provides
correct electronic structures and anisotropies but the size of the
inter-ionic exchange parameters, and therefore spin-excitations, are overestimated
within DFT~\cite{molmag}, due to slight delocalization of the $d$-electrons.
A class of  Ni$_4$O$_4$ molecular magnets has proven to be particularly problematic within standard DFT approaches~\cite{park,cheng}.
Inclusion of a Hubbard U treatment, an approximation to SIC, is needed to correctly predict the spin-ordering~\cite{cheng}.

In the original formulation of the problem, for a given approximation to
the exchange correlation functional, $E_{xc}^{approx}[n_\uparrow,n_\downarrow]$, Perdew and Zunger~\cite{R1}
suggested appending the following term to the DFT functional:
\begin{equation}
E_{xc}^{PZ-SIC} =
-\Sigma_{\alpha,\sigma} \{ U[n_{\alpha,\sigma}]+
E_{xc}^{approx} [n_{\alpha,\sigma},0]\}
\end{equation}
              
In the above equation, the orbitals  $\{\phi_{\alpha\sigma}\}$ are used to define
orbital densities according to: $ n_{\alpha \sigma}({\bf r})= |\phi_{\alpha\sigma}(\bf r)|^2$.
  The terms
$U[n_{\alpha,\sigma}]$ and $E_{xc}^{approx} [n_{\alpha,\sigma},0]$ are the exact self-coulomb and approximate
self exchange-correlation energies, respectively. The PZ paper recognized that
this formulation led to a definition for the
energy functional that did not transform like the density and posited that localized
orbitals similar to those proposed
by Edmiston and Ruedenberg~\cite{R36P} might be the most appropriate set of orbitals for defining
the SIC~\cite{R1}. Shortly
thereafter, Lin's Wisconsin SIC group followed up on this suggestion and introduced the concept of
localized and canonical orbitals in self-interaction corrected theories~\cite{heatonra,R32,R33,R33P}.
These papers showed that to ensure a Hermitian Lagrange multiplier matrix, the
orbitals used for constructing the SIC energy must satisfy the $O(N^2)$ localization equations given by
\begin{equation}
<\phi_{i\sigma}|V^{SIC}_{i\sigma}-V^{SIC}_{j\sigma}|\phi_{j\sigma}>=0,
\end{equation}
with $V^{SIC}_{i\sigma}$ the partial functional derivative of Eq.~(1) with respect to the orbital density $n_{i\sigma}$. The Jacobi-like
approach~\cite{R33} to solving these equations also depended on $O(N^2)$ Jacobi updates.
The localized orbitals obtained from these equations were found to be topologically similar to $sp^3$ hybrids in atoms, alternative
energy-localized orbitals in molecules~\cite{R36P} and Wannier functions in solids~\cite{heatonra}. In addition, they
satisfied an explicitly local Schr{\"o}dinger-like equation that
was coupled together by off-diagonal Lagrange multipliers. On the other hand, the canonical orbitals were topologically similar to
the DFT KS orbitals.

\begin{figure}[h!]
\begin{center}
\psfig{file=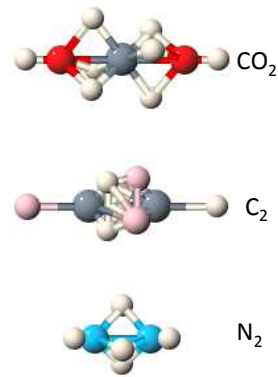,height=2.5in,clip=true}
\end{center}
\caption{
Valence Fermi Orbital Centroids (FOC) are superimposed over
molecular equilibrium geometries for $CO_2$,  $C_2$, and $N_2$.  In all cases
the FOCs form vertices on distorted tetrahedra.
}
\label{fig2}
\end{figure}
 The one-particle density matrices, $\rho_\sigma({\bf r},{\bf r'})$ and
densities $n_\sigma({\bf r})=\rho_\sigma({\bf r},{\bf r})$, arising from single determinants or products of
single determinants  are invariant under unitary transformation
of the occupied orbitals of each spin, and the value of the many-electron wavefunction can only
change by a phase factor under such transformations. The resulting energy expectation value
from such wavefunctions
is also invariant to unitary transformations within the space of orbitals used for
constructing wavefunctions, and transforms like the one-particle density matrices. While
DFT also exhibits these symmetries, in SIC-DFT
there are many possible choices for a unitary
transformation from Kohn-Sham to SIC orbitals (real or complex~\cite{simon},
satisfying energy-minimizing or other conditions). The standard ways
to find SIC orbitals are computationally demanding, especially for systems
of many atoms, and results have raised questions about size
extensivity~\cite{R28} in SIC-DFT~\cite{simon2}.
While the uncorrected functionals do not have problems with size extensivity, the SIC functionals fail to be so if the SIC
orbitals delocalize over distant atomic sites.  
These problems are solved in this Letter.
                                         
Here, a modification of the original formulation of the Perdew-Zunger self-interaction
correction is introduced. For notational simplicity, we consider the large number of molecular and crystalline
systems that can, in principle, be described by real KS orbitals and that have an integer
number of spin up and spin down electrons.
This formulation leads to a size-consistent  SIC {\em spin-density} functional that is invariant to unitary
transformations in the occupied orbital space and leads to a long-range effective potential
that scales as $-1/r$. Rather than allowing the SIC localized orbitals to be {\em any} unitary transformation within
the occupied orbital space, we introduce a constraint
that the orbitals used for constructing Eq.~(1) must be explicitly dependent on a quantity that
is itself unitarily invariant. The Fermi orbital (FO)~\cite{FO1,FO3} is a specific example of such a quantity.
Given a trial set of KS orbitals, the FO ($F_{i\sigma}$) is defined at any point in space,
${\bf a_{i\sigma}}$,  according to:
\begin{eqnarray}
F_{i\sigma}(\bf r)=\frac{\rho_\sigma({\bf a}_{i\sigma},{\bf r}) }{\sqrt{\rho_\sigma({\bf a}_{i\sigma})}}, \\
F_{i\sigma}({\bf r})=
\frac{\Sigma_{\alpha}  \psi^{*}_{\alpha\sigma}
({\bf a}_{i\sigma})
\psi_{\alpha\sigma}(\bf r)}{\sqrt{\{ \Sigma_{\alpha}|\psi_{\alpha\sigma} \bf(a_{i\sigma})|^2\}}}
\equiv\Sigma_{\alpha}T_{i\alpha}^{\sigma}
\psi_{\alpha\sigma}({\bf r}).
\end{eqnarray}
In other words, the FO is simply the
ratio of the one-particle spin-density matrix to the square root of the spin density and is ultimately a simple transformation
of the KS orbitals.
It is easy to verify that Wannier functions are a sub-class of Fermi Orbitals and
a few more  comments illustrate their physical and chemical nature.
At ${\bf r}={\bf a_{i\sigma}}$, the value
of the absolute square of the FO is identically equal to the total spin density at
${\bf r}={\bf a_{i\sigma}}$. Further, the FO associated with any position, ${\bf a_{i\sigma}}$, in space is normalized
to unity. For special sets of points, the FOs can be immediately orthogonal (e.g. Wannier Functions)  but
for general sets of points, they are not orthogonal.
Second, the absolute square of the FO is minus the
exchange hole density at ${\bf r}$ around an electron at ${\bf a_{i\sigma}}$.
The exact exchange energy of a single Slater determinant is simply
\begin{equation}
E^x = -\frac{1}{2}\Sigma_\sigma
\int d^3r \int d^3a \frac{
|\Sigma_\alpha \psi^*_{\alpha\sigma}({\bf r})
\psi_{\alpha\sigma}({\bf a})|^2}{|{\bf r - a}|}.
\end{equation}
In Ref.~\cite{exchange-hole},
it was noted that the minus the exchange-hole density had characteristics similar to a
single-orbital density but a simple-to-use
local expression for replacing the non-local Fock operator remained difficult to obtain.

Because the FO accounts for all the spin density at a given point in
space it is expected to be a rather localized function. So, in order to incorporate self-interaction corrections for
the states with spin $\sigma$ in a system that contains a total of $N_\uparrow + N_\downarrow$ electrons,
the Perdew-Zunger formulation of the
self-interaction correction can be slightly constrained by invoking the following strategy for each spin:
\begin{enumerate}
\item  For a trial set of KS orbitals $\{\psi_{\alpha \sigma}\}$ find N$_\sigma$ centroids
$\{{\bf a}_{1\sigma}, {\bf a}_{2\sigma},...,{\bf a}_{N_\sigma\sigma}\}$which provide a set of N$_\sigma$ normalized linearly
independent, but not orthogonal
FO $\{F_{1\sigma},F_{2\sigma}...F_{N_\sigma \sigma}\}$
which, from Eq.~(4), will always
lie in the space spanned by the KS orbitals.
\item Use L{\"o}wdin's method of symmetric orthonormalization~\cite{R33,Low} to transform the set of FO to a set
of localized orthonormal orbitals
$\{\phi_{1\sigma},\phi_{2\sigma}...,\phi_{N_\sigma \sigma}\}$ that are a unitary transformation on the KS orbitals.
\item Minimize the energy as a function of the KS orbitals and the classical
centroids of the FOs.  The optimization of the KS orbitals can follow any of the approaches
described in Ref.~\cite{RLT} and references therein. The Fermi-orbital centroids (FOCs) can be determined using methods that
are commonly used for optimizing molecular geometries.
\end{enumerate}
This approach bypasses solution of the localization equations in entirety at the lesser expense of the search
for quasi-transferable and chemically appealing semi-classical FOCs (Fig. 1) on which the FO and unitary transformation depend.
                                                                                                                               
Compared to the localization equations~\cite{R33},
the FO-formalism adds a constraint that prevents consideration of all unitary transformations and especially discourages the use of
the Kohn-Sham orbitals such as
Bloch Functions, molecular orbitals or orbital-angular momentum states in atoms.
A simple molecular case that illustrates this is the He$_2$ molecule. To obtain $\sigma_g$ and $\sigma_u$
FO, it would be necessary to find a point where the value of the nodeless $\sigma_g$ state is zero and the value of the $\sigma_u$ state
is non-zero, which is impossible.
                                 
{\bf Methodology and Scaling:}
Solution of the KS equations is usually $N^3$, unless special linear-scaling methods are employed.
The FO formalism reduces the number of variational parameters
from $O(N^2)$ to $O(3N)$ and in this regard should scale faster than methods based upon solution of Eq.~(2). Our analysis is that, for
systems with a gap, an algorithm based upon direct iterative updates of the FOs (rather than the KS orbitals), would scale as favorably as
DFT ($O(N)-O(N^2)$) in the many-atom limit.
Also, compared to the methods of Ref.~\cite{R32,R33,R33P} the energy determined from the FO formalism should be higher as a
result of the constraint that FSIC is explicitly dependent on the density.
The calculations discussed here used a modified version of NRLMOL~\cite{nrlmol,ppp}.
Large Gaussian-orbital basis sets~\cite{nrlmol} were used to  represent the electron
wavefunctions.
\begin{center}
\begin{table}[htp]
\begin{tabular}{|l|r|r|r|r|r|}
Mol. & LSD  & GGA & SIC & FSIC& Expt \\
     & PW92 & PBE & PZ  & PW92&   \\
\hline
$H_2$     & 4.91     & 4.51     &4.97   &  4.97 & 4.77 \\
$Li_2$    & 1.03     & 1.06     &1.04   &  1.02 & 1.06 \\
$LiF$     & 6.75     & 6.01     &6.17   &  5.61 & 6.03 \\
$N_2$     &11.58     & 10.49    &10.89  &  9.80 &9.84 \\
$O_2$     & 7.62     & 6.30     &5.77   &  4.80 & 5.12 \\
$CO$      & 12.94    & 11.65   &12.02  &  11.00 & 11.32 \\
$CO_2$    &20.57     &18.16     &18.29  & 16.88&17.00 \\
$CH_4$    &20.06     &18.24     &20.25  & 20.23&18.21 \\
$NH_3$    &14.56  &13.05  &14.21  & 14.24 &12.88\\
$H_2O$    &11.64     &10.27     &10.68  & 10.71&10.10 \\
$C_2H_2$  & 19.93    & 18.01    &19.81  & 18.93  &17.52   \\
$C_2$     & 7.23     & 6.22     &          &  5.10&6.31 \\
\hline
\end{tabular}
\caption{Atomization energies (eV) of molecules as determined from FSIC-PW92 (this work). LSD-PW92[3],
GGA-PBE[4] and experimental results
are from Refs. [21-22,38]. SIC-PZ results are from Ref. [32].}
\end{table}
\end{center}
            
{\bf Applications to Atoms:} As a first test we have performed FSIC calculations on the six lightest
closed shell-atoms ($He$, $Be$, $Ne$, $Mg$, $Ar$, $Ca$). In all cases, the FO procedure provides localized orbitals that
resemble the $sp^3$ hybrids and $1s-$core orbitals identified in Ref.~\cite{R33P}.
                                                                                  
\begin{table}[bh]
\begin{tabular}{|rrrrrrr|}
\hline
State &  HF & SIC    & FSIC & LSD& Expt & CASSCF \\
      &     & LSDX   & PW92 & PW92& Expt &  \\
\hline
$1\sigma_g$ (au) &  -15.709  & -15.639      &  -15.68          &    -13.966  &   &       \\
$1\sigma_u$ (au)&  -15.706   & -15.637      &  -15.67          &    -13.965  &      &    \\
$2\sigma_g$ (au)&   -1.525  & -1.428      &  -1.371          &    -1.0422  &   &       \\
$2\sigma_u$ (au)&   -0.775 & -0.745      &   -0.788          &    -0.492    &    &     \\
$1\pi_u$    (au)&   -0.62 &-0.639      &   -0.687         &      -0.438  & &    \\
$3\sigma_g$ (au) &   -0.631 & -0.600      &   -0.658          &      -0.388  & -0.573 &   \\
$D_e$ (eV)  &            &   7.26         &   9.8 eV      &    11.58        &     9.80     &  9.84      \\
\hline
\end{tabular}
             
\caption{The nitrogen molecule. Eigenvalues (Hartree) and
dissociation or atomization energy ($D_e$ in eV) for Hartree-Fock (HF), self-interaction corrected
exchange-only (SIC-LSDX)[29,31], FSIC (this work), LSD-PW92 (this work), experiment and CASSCF[39].}
\end{table}
           
{\bf Applications to molecules:}
In Table I we compare calculated atomization energies from LSD-PW92~\cite{PW92}, GGA-PBE~\cite{pbe},
SIC-PZ~\cite{simon}, FSIC-PW92 and experiment. The FSIC-PW92
provides significant improvements over LSD as compared to experiment and does as well or better than PBE-GGA in some
cases. The results in Table I are encouraging and suggest that this class of self-interaction corrected functionals
may provide higher accuracy.  To provide greater technical detail and compare to other calculations
we discuss two cases in a bit more detail.
                                          
The strongly covalent singlet N$_2$ molecule, $1\sigma_g^2 1\sigma_u^2 2\sigma_g^2 2\sigma_u^2 1\pi_u^4 3\sigma_g^2$,
which dissociates into an open-shell singlet with
three unpaired $2p$ electrons per atom is challenging to represent continuously as a broken-symmetry single determinant.
Within the FO-method, the atomization energy for the N$_2$ molecule ($R_e^{PW92}= 2.071 au$) is determined by taking
the difference in total energy between the spin unpolarized molecule and the spin polarized atoms. Using FSIC-PW92, we find
an atomization energy (D$_e$) of 9.80 eV which compares well to the experimental atomization energy (9.84 eV) and to
high-accuracy CASSCF results (9.85 eV) of Li Manni {\em et al}~\cite{manni}.
For comparison, the LSD-PW92 energy functional gives an atomization energy of 11.54 eV at this bondlength and the GGA-PBE energy
functional gives an atomization energy of 10.54 eV. In Fig.~1, the valence FOCs are shown pictorially.
In accord with Refs.~\cite{R32,simon}, we have ascertained that the FO-formalism creates doubly
occupied $1s$ FO on both atoms, $1s_{A_+/A_-}=\{1\sigma'_g \pm 1\sigma'_u\}/\sqrt{2}$,
lone-pair states on the exterior of the molecule, $2sp_{A+/A_-}=\{3\sigma'_g \pm 2\sigma'_u\}/\sqrt{2}$,
and three bond-centered banana orbitals
(e.g. $ \phi_n=\frac{1}{\sqrt{3}}[2\sigma_g'-\sqrt{2}\{\cos(\frac{2n\pi}{3})\pi_{ux} + \sin(\frac{2n\pi}{3}) \pi_{uy}\}]$,
with $n=-1,0,+1$ ).
As in Ref.~\cite{R32}, the primes indicate that KS molecular orbitals of the same symmetry are mixed together by a
unitary transformation within each irreducible representation to minimize Eq.~(1).
For example the $\{2\sigma_g',3\sigma_g'\}$ are not perfect eigenstates. Instead they are determined by a nearly
diagonal unitary mixture of the $\{2\sigma_g,3\sigma_g\}$ KS eigenstates.
In Table II, we compare eigenvalues as calculated from HF, an earlier SIC-LSDX calculation~\cite{R33}, and experiment.
The results show that the eigenvalues move to significantly lower energies in accordance with previously identified
trends.  In accord with the results of Ref.~\cite{R33P}, the
FOs for the separated atom are $2sp^3$ hybrids for the majority spin.

  Within FSIC-PW92 we find that the atomization energy
for  methane is 20.23 eV, in good agreement with the results of Kl{\"u}pfel {\em et al}~\cite{simon} ($R_e^{PW92}(C-H)=2.074$ Bohr).
However, this is not an improvement over PW92 (20.06 eV).
For comparison, the GGA-PBE and experimental atomization energy are 18.24 and 18.21, eV respectively.  For
this case there is a $1s$ core FOC on top of the C atom and there are four equivalent
$2sp^3$ FO that are composed of C $2s$, $2p$, and H $1s$ character.
The minimal energy is found when the FOC is 1.82 Bohr from the $C$ atom along each $C-H$ bond.
The FSIC-PW92 energy is weakly dependent on the FOC and changes by only 0.1 eV when FOCs move 1.2-2.2 Bohr, along their CH bonds,
from the $C$ atom.
However, it may not be surprising that the optimal FOC
is in very good agreement with the position of maximum density commonly identified by x-ray spectra.
In this regard, it may be useful to analyze x-ray data for $N_2$ molecules to understand whether either the $1s$-FOC or the centroid
of its orthonormal child,
$<\phi_{1s\sigma}|{\bf r}|\phi_{1s\sigma}>$, compare with positions of maximal density identified by x-ray analysis.
Within FSIC-PW92, the energy of the hydrogen atom is exact and the FOs for the isolated C atom are qualitatively similar to those identified in Ref.~\cite{R33P}.
                                                                                                                                                                 
To summarize, we present and test a simplified computationally efficient and systematic theoretical 
framework for incorporating self-interaction corrections into
the density-functional approximation. 
Compared to the accuracy of GGA-PBE the results for FSIC-PW92 are mixed but still encouraging.
The combination of an asymptotically correct long-range potential and explicit unitary invariance
offered by the FO may allow for future meta- and/or hyper- gradient corrected
formalisms that are constructed in an {\em a priori} self-interaction-free form.
The FOs obtained here are topologically similar to those obtained from the localization-equations used earlier~\cite{R33P}.
However, the unitary transformations that can be constructed from the FOs are constrained and, relative to the original
formulation, retains the symmetries and size extensivity that are present in density-functional theory.
                                                                
{\bf Acknowledgements:} MRP thanks Dr. R.A. Heaton who first suggested the possibility of using a FO in
the PZ SIC functional. JPP acknowledges support from NSF Grant No. DMR-1305135.


\begin{references}
\bibitem{R2} P. Hohenberg and W. Kohn, Phys. Rev. {\bf 136}, B864 (1964).
\bibitem{R3} W. Kohn and L.J. Sham, Phys. Rev. {\bf 140}, A1133 (1965).
\bibitem{PW92}J. P. Perdew, J. A. Chevary, S. H. Vosko, K. A. Jackson, M. R. Pederson, D. J. Singh, and C. Fiolhais,
Phys. Rev. B {\bf 46}, 6671 (1992).
\bibitem{pbe}J. P. Perdew, K. Burke, and M. Ernzerhof, Phys. Rev. Lett. {\bf 77}, 3865  (1996).
\bibitem{R1} J.P. Perdew and  A. Zunger, Phys. Rev. B {\bf 23}, 5048 (1981).
\bibitem{R28} J.P. Perdew, ``Size-Consistency, Self-Interaction Correction, and
Derivative Discontinuity'', in {\em Density Functional Theory of Many-Electron
Systems}, edited by S.B. Trickey, Advances in Quantum Chemistry
{\bf 21}, 113 (1990).
\bibitem{RLT}M.R. Pederson and J.P. Perdew, ``Self-Interaction Correction in Density
Functional Theory: The Road Less Traveled", $\Psi_K$ Newsletter Scientific Highlight
of the Month, February (2012).
\bibitem{aron}A. J. Cohen, P. Mori-Sanchez, and W. Yang, Science {\bf 321}, 5890 (2008).
\bibitem{dipole1}N. I. Hammer, R. J. Hinde, R. N. Compton, K. Diri, K. D. Jordan, D. Radisic, S. T. Stokes and K. H. Bowen, J. Chem. Phys.
{\bf 120}, 685 (2004).
\bibitem{dipolebound}K.D. Jordan and F. Wang, Ann. Rev. of Physical Chem. {\bf 54}, 367-396 (2003).
\bibitem{mott}N.F. Mott and R.W. Gurney, {\em Electronic Processes in Ionic Crystals} (Dover, New York, 1965 pp 80 and 114.
\bibitem{klein}M.R. Pederson and B.M. Klein, Phys. Rev. B {\bf 37} 10319 (1988).
\bibitem{dederich} P.H. Dederichs, S. Bl{\"ugel}, R. Zeller and H. Akai, Phys. Rev. Lett. {\bf 53} (1984).
\bibitem{baruah}M. Olguin, T. Baruah, and R. Zope, J. Chem. Phys.  {\bf 138} 074306 (2013).
\bibitem{dissociation}A. Ruzsinszky, J. P. Perdew, G.I. Csonka, O.A. Vydrov and
G.E. Scuseria J. Chem. Phys. {\bf 125} 194112 (2006).
\bibitem{molmag}A. Postnikov, J. Kortus, and M.R. Pederson, Phys. Stat. Solidi (b) {\bf 243}, 2533 (2006).
\bibitem{dilemma}J.P. Perdew, A. Savin, and K. Burke, Phys. Rev. A {\bf 51}, 4531 (1995).
\bibitem{sanvito}  D. Pemmaraju, T. Archer, D. Sanchez-Portal, S. Sanvito, Phys. Rev. B {\bf 75} 045101 (2007).
\bibitem{tddft}E. Runge and E.K.U. Gross, Phys. Rev. Lett. {\bf 52}, 997 (1984).
\bibitem{hoffmann}D. Hoffmann and S. K{\"u}mmel, J. Chem. Phys. {\bf 137}, 064117 (2012).
\bibitem{ernzerhof}M. Ernzerhof and G.E. Scuseria, J. Chem. Phys. {\bf 110}, 5029 (1999).
\bibitem{vydrov} O.A. Vydrov, and G.E. Scuseria, J. Chem. Phys. {\bf 121} 8187 (2004).
\bibitem{janak}J.F. Janak, Phys. Rev. B {\bf 18}, 7165 (1978).
\bibitem{temmerman}A. Svane, V. Kanchana, G. Vaitheeswaran, G. Santi, W.M. Temmerman, Z. Szotek, P. Strange,
 and L. Petit, Phys. Rev. B {\bf 69}, 054427 (2004).
\bibitem{park}K. Park, E.-C. Yang, and D.N. Hendrickson, J. Appl.
Phys. {\bf 97}, 10M522 (2005).
\bibitem{cheng}C. Cao, S. Hill, and H.P. Cheng,
Phys. Rev. Lett. {\bf 100}, 167206-4 (2008).
\bibitem{R36P}C. Edmiston and K. Ruedenberg, Rev. Mod. Phys. {\bf 35}, 457 (1963).
\bibitem{heatonra}R.A. Heaton, J.G. Harrison and C.C. Lin, Phys. Rev. B {\bf 28}, 5992 (1983).
\bibitem{R32} M.R. Pederson, R.A. Heaton, and C.C. Lin, J. Chem. Phys. {\bf 80},
1972 (1984).
\bibitem{R33} M.R. Pederson, R.A. Heaton, and C.C. Lin, J. Chem. Phys. {\bf 82},
2688 (1985).
\bibitem{R33P}M.R. Pederson and C.C. Lin, J. Chem. Phys. {\bf 88}, 1807 (1988).
\bibitem{simon}S. Kl{\"u}pfel, P. Kl{\"u}pfel and H. J\'onsson, J. Chem. Phys. {\bf 137}, 124102 (2012).
\bibitem{simon2} S. Kl{\"u}pfel, P. Kl{\"u}pfel, and H.J\'onsson, Phys. Rev. A
{\bf 84}, 050501(R) (2011). Fig. 1 of this paper shows positive
SIC corrections to the GGA-PBE total energies of atoms
heavier than N, from real SIC orbitals. P. Kl{\"u}pfel plausibly
concluded from this that, in a sufficiently-stretched diatomic
molecule made of two identical heavy atoms, the real SIC orbitals
would delocalize, leading to a failure of size extensivity (public
discussion at the CECAM Workshop on Self-Interaction Correction:
State of the Art and New Directions, Chester, England, 2011). 
\bibitem{FO1} W.L. Luken and D.N. Beratan, 
Theor. Chim. Acta, {\bf 61}, 265-281 (1982).
\bibitem{FO3} W.L. Luken and J.C. Culberson,
Theor. Chim.  Acta, {\bf 66} 279-283 (1984).
\bibitem{exchange-hole}J. P. Perdew, ``Nonlocal Density Functionals for Exchange and Correlation" in
{\em Density Functional Theory of Molecules, Clusters and Solids}, Ed. by D. E. Ellis (Kluwer Academic Publishers, 1995).
\bibitem{Low}P.O L{\"o}wdin, Rev. Mod. Phys. {\bf 34}, 520 (1962).
\bibitem{nrlmol} M.R.  Pederson, DV Porezag, J. Kortus and DC Patton, Phys. Stat. Solidi B {\bf 217},  197 (2000).
\bibitem{ppp}D.C. Patton, D.V. Porezag and M.R. Pederson, Phys. Rev. B {\bf 55} 7454 (1999).
\bibitem{manni}G. Li Manni, D. Ma, F. Aquilante, J. Olsen and L. Gagliardi, J. Chem. Theory and Comput. {\bf 9} 3375 (2013).
\end{references}
\end{document}